\documentclass[lettersize,journal]{IEEEtran}
\usepackage{amsmath,amsfonts}
\usepackage{algorithmic}
\usepackage{algorithm}
\usepackage{array}
\usepackage[caption=false,font=normalsize,labelfont=sf,textfont=sf]{subfig}
\usepackage{textcomp}
\usepackage{stfloats}
\usepackage{url}
\usepackage{verbatim}
\usepackage{graphicx}
\usepackage{cite}
\usepackage{amsthm}
\usepackage{wrapfig}

\theoremstyle{definition}

\begin{document}

\title{Federated Learning with Blockchain-Enhanced Machine Unlearning: A Trustworthy Approach}
\author{Xuhan Zuo, Minghao Wang, Tianqing Zhu$^*$, Lefeng Zhang, Shui Yu,~\IEEEmembership{Fellow,~IEEE,} Wanlei Zhou,~\IEEEmembership{Senior Membership,~IEEE}
\thanks{$^*$Tianqing Zhu is the corresponding author with Faculty of Data Science, City University of Macau, Macao (E-mail: tqzhu@cityu.edu.mo)}
\thanks{Xuhan Zuo and Shui Yu are with School of Computer Science, University of Technology Sydney, Ultimo 2007, Australia (E-mail: Xuhan.Zuo-1@student.uts.edu.au; Shui.Yu@uts.edu.au)}
\thanks{Minghao Wang, Lefeng Zhang and Wanlei Zhou are with the Faculty of Data Science, City University of Macau, Macao (E-mail: sydminghao@gmail.com; lfzhang@cityu.edu.mo; wlzhou@cityu.edu.mo)}
}



\maketitle

\begin{abstract}

With the growing need to comply with privacy regulations and respond to user data deletion requests, integrating machine unlearning into IoT-based federated learning has become imperative. Traditional unlearning methods, however, often lack verifiable mechanisms, leading to challenges in establishing trust. This paper delves into the innovative integration of blockchain technology with federated learning to surmount these obstacles. Blockchain fortifies the unlearning process through its inherent qualities of immutability, transparency, and robust security. It facilitates verifiable certification, harmonizes security with privacy, and sustains system efficiency. We introduce a framework that melds blockchain with federated learning, thereby ensuring an immutable record of unlearning requests and actions. This strategy not only bolsters the trustworthiness and integrity of the federated learning model but also adeptly addresses efficiency and security challenges typical in IoT environments. Our key contributions encompass a certification mechanism for the unlearning process, the enhancement of data security and privacy, and the optimization of data management to ensure system responsiveness in IoT scenarios.

\end{abstract}


\section{Introduction}

\IEEEPARstart{T}{he} introduction of the Internet of Things (IoT) has initiated a new era in service computing and collaborates with the field of federated learning, an approach to machine learning that allows collaborative model building among multiple clients~\cite{khan2021federated}. 
However, according to recent regulations like the GDPR \cite{regulation2018general} and CCPA \cite{bukaty2019california}, users have the right to request the deletion of their data. Consequently, there emerged a new technology called machine unlearning, that integrate into federated learning to meet those regulations~\cite{xu2023machine}. The need for machine unlearning stems from the imperative to systematically remove specific data from trained models in IoT environments. Currently, there are several methods to achieve machine unlearning in federated learning~\cite{bourtoule2021machine}\cite{baumhauer2022machine}. 

Unfortunately, 
existing machine unlearning techniques suffer from one critical issue: when they have deployed in servicing computing: the model owner cannot provide users with evidence that their data has truly been unlearned. 
In other words, users must blindly believe that the server has executed the unlearning oracle to remove the influence of their data, with little opportunity to verify it \cite{eisenhofer2022verifiable}. This makes the entire unlearning process untrustworthy and may lead to concerns about privacy. An untrustworthy service providers may pretend to have performed the unlearning operation but actually take no action to avoid high computational costs or utility decreasing~\cite{ghorbani2019data}.

Blockchain technology, renowned for its immutability, transparency, and robust security, does more than just establish a foundation of trust for digital transactions and record-keeping. It also extends this trust paradigm to machine unlearning\cite{wang2023differentially}. This integration leverages the strengths of blockchain to address trust issues, offering several key advantages: 
First, blockchain guarantees that each step in the unlearning process is immutably recorded, ensuring a verifiable and transparent certification of data removal actions. This level of transparency is crucial for auditability, allowing stakeholders to easily trace and verify all unlearning actions, thereby assuring the integrity of the process.
Second, blockchain's robust security framework is instrumental in preventing unauthorized alterations, a vital aspect for maintaining reliable certification. Finally, its decentralized nature spreads the trust mechanism across multiple nodes, effectively eliminating single points of failure and bolstering the system's resilience in certifying unlearning actions. The convergence of these benefits significantly bolsters the trustworthiness of the unlearning process in federated learning environments.

The blockchain layer functions as a decentralized ledger, providing an audit trail that confirms the accurate execution of unlearning requests. This design significantly increases the trustworthiness and integrity of the federated learning model.
If we introduce a trustworthy machine unlearning framework that is based on the combination of blockchain and federated learning, we will have a certified machine unlearning technology. 


However, there are several challenges when we integrate the blockchain with machine unlearning, 

\begin{itemize}
\item \textbf{Certification:} Certification presents a complex issue, as it necessitates the system to authenticate an unlearning request in a decentralized environment.
\item \textbf{Security and Privacy:} Balancing security and privacy is essential; although blockchain improves the transparency of the machine unlearning process, it potentially risks compromising the confidentiality of sensitive data. This is because every user in the blockchain network might have access to view the content.
\item \textbf{Efficiency:} Efficiency remains a key concern. The integration of blockchain could potentially introduce latency or computational overhead, which needs to be carefully managed to preserve agility and responsiveness, especially in the context of IoT.
\end{itemize}

This paper tackles these challenges by integrating of blockchain in federated learning guarantees an unalterable record of unlearning requests and actions, thereby enhancing certification, efficiency, and establishing a robust security and privacy framework. 
For certification, our framework employs smart contracts to automate and streamline the verification and validation of unlearning requests, ensuring their authenticity and accurate processing. Regarding security and privacy, we integrate differential privacy mechanisms into the smart contracts. This integration adds an extra layer of data protection, safeguarding confidentiality while maintaining transparent unlearning processes. Efficiency is enhanced by optimizing the storage of federated learning training process data. This adjustment minimizes computational overhead, ensuring that blockchain technology integration does not compromise the system's performance too much, especially in time-sensitive IoT applications.

Our contributions are summarized as follows:

\begin{itemize}
    \item We proposed a novel framework that integrates blockchain technology with federated learning specifically for machine unlearning, providing a verifiable certification of the unlearning process in decentralized environments. This certification authenticates and validates each unlearning request, offering crucial proof within IoT contexts.
    \item We enhanced the security and privacy of machine unlearning in federated learning systems. Utilizing blockchain's transparent ledger and smart contract capabilities, it establishes a secure method to protect data security and privacy.
    \item We optimized the data management process in machine unlearning by combining a blockchain network with federated learning in an IoT scenario. This approach efficiently stores training process data in federated learning, ensuring system agility and responsiveness when an unlearning request is initiated in the blockchain network.
\end{itemize}

The paper is structured as follows. Section II covers related work, providing a comprehensive review of the existing literature on Federated Learning, Machine Unlearning, and Blockchain technology. Section III discusses the foundational concepts and technical background of these areas. Section IV outlines the specific challenges and objectives that our research addresses and presents our proposed system conceptual model. Section V describes our novel framework, its architecture, functionalities, and the integration of blockchain with federated learning. Section VI evaluates the privacy and security aspects of our system. Section VII presents empirical assessments, and Section VIII summarizes our findings and suggests directions for future research.

\section{Related work}

\subsection{Federated Learning with Machine Unlearning}
The integration of federated learning and machine unlearning is a subject of increasing interest in current research. However, several studies do not adequately address the issue of trustworthiness. For instance, Liu et al.\cite{liu2022right} delve into this area by proposing a rapid retraining method for efficient data removal in federated learning (FL), crucial for adhering to the 'right to be forgotten.' This approach highlights the challenges of implementing unlearning in decentralized systems like FL, where data remains local to participants. Our proposed method builds upon this by incorporating blockchain technology into the FL framework for machine unlearning. This not only maintains FL efficiency but also adds a new level of transparency and accountability through the immutable ledger of blockchain, providing a verifiable and tamper-proof record of unlearning activities, crucial for IoT environments.

Wu et al.\cite{wu2022federated} address the "Right to be Forgotten" in federated learning with a focus on federated unlearning. They outline efficient techniques for removing data influences from models, such as reverse SGA and EWC. While providing a solid unlearning framework, our method further enhances the process by integrating blockchain technology, ensuring verifiability and trustworthiness. This is particularly important for the sensitive nature of IoT applications, streamlining efficiency while fortifying the integrity of the unlearning process.

Wang et al.\cite{wang2023federated} focus on enforcing "the right to be forgotten" in federated learning systems, identifying the issue of information leakage due to model discrepancies before and after unlearning. Their work reviews existing studies proposes new taxonomies for unlearning algorithms, and highlights vulnerabilities to inference attacks. In contrast, our method integrates blockchain technology with federated unlearning, enhancing both the efficiency and security of the process. It ensures an immutable, transparent record of unlearning activities, thereby minimizing information leakage and unauthorized data access.

Wang et al.\cite{wang2022federated} introduce a method for selective unlearning in federated learning (FL) using channel pruning based on Term Frequency-Inverse Document Frequency (TF-IDF) for CNN classification models. This accelerates unlearning while maintaining model accuracy. Our approach, however, enriches the FL unlearning process by integrating blockchain technology, adding a layer of transparency and trust, essential for security and regulatory compliance in IoT applications.

Zhang et al.\cite{zhang2023fedrecovery} present "FedRecovery," an efficient algorithm for machine unlearning in federated learning, overcoming the limitations of traditional retraining-based methods. It leverages differential privacy for indistinguishability in complex tasks like neural networks. Our method, by integrating blockchain technology, adds an essential layer of transparency and trustworthiness to the unlearning process, meeting the high security and compliance demands in IoT applications.

Liu et al.\cite{liu2021federaser} propose "FedEraser," an approach in federated learning that efficiently removes specific training data from FL models, outperforming traditional methods in speed. It uses historical parameter updates for rapid model reconstruction. Our method enhances this by integrating blockchain technology into the federated unlearning process, improving efficiency and significantly boosting transparency and trustworthiness, vital for handling sensitive data in IoT applications.

\subsection{Blockchain with Federated Learning}

The integration of blockchain in federated learning has been explored in various contexts, but the specific question of machine unlearning within this framework appears to be largely unaddressed. For instance:

Nguyen et al.\cite{nguyen2021federated} discuss FLchain, a novel integration of mobile-edge computing (MEC), federated learning (FL), and blockchain, which enhances privacy and security in decentralized AI training. This concept emphasizes collaborative training across multiple mobile devices without data exposure. In comparison, our research uniquely focuses on machine unlearning within this FLchain framework, a topic scarcely addressed in existing studies. Our approach uses blockchain not only for secure and private data management but also to enable trustworthy and verifiable unlearning processes in FL systems, advancing responsible AI practices in MEC environments.

Pokhrel et al.\cite{pokhrel2020federated} introduce an autonomous blockchain-based federated learning system for vehicular networks, emphasizing decentralized, privacy-aware, and efficient communication. It combines on-vehicle machine learning and blockchain consensus mechanisms. However, our research augments this domain by incorporating machine unlearning into the blockchain-based federated learning (BFL) system, an aspect overlooked in their study. Our integration enhances both data processing efficiency and compliance with data regulations, making it highly relevant in dynamic environments like vehicular networks.

Li et al.\cite{li2020blockchain} propose the Block-chain-based Federated Learning with Committee consensus (BFLC) framework, aiming to secure federated learning against malicious attacks. BFLC leverages blockchain for decentralized model storage and local model updates. Our research enriches this framework by integrating machine unlearning, a feature not included in BFLC. This addition not only addresses security concerns but also enables verifiable data removal, ensuring compliance in rapidly changing data environments.

Lu et al.\cite{lu2019blockchain} present a blockchain-based architecture for secure data sharing in the industrial Internet of Things, focusing on privacy in wireless networks. They integrate privacy-preserved federated learning into the blockchain consensus process. Our research extends this by embedding machine unlearning into the blockchain-enhanced federated learning framework, addressing the need for dynamic data management and adherence to privacy regulations in the industrial IoT.

Shayan et al.\cite{shayan2020biscotti} present Biscotti, a decentralized peer-to-peer approach for multi-party machine learning, using blockchain and cryptographic methods to enhance privacy and security, while addressing the limitations of federated learning. Our research further advances this concept by integrating machine unlearning capabilities into the decentralized learning framework, maintaining Biscotti's robust privacy and security features while adding the ability to manage and remove data responsibly, ensuring compliance with evolving privacy standards.

In summary, while various studies have explored blockchain's application in federated learning for enhancing security and privacy, our research takes a novel step by integrating machine unlearning into this paradigm. This integration addresses a crucial gap in the current literature and positions our approach as a pioneering solution in the realm of secure, dynamic, and responsible federated learning systems.

\section{Preliminary}

\subsection{Federated Learning}

Federated Learning (FL) is a machine learning paradigm where a model is trained across multiple decentralized devices (or clients) holding local data, without the need for exchanging the data itself. This approach is key for preserving data privacy and minimizing the necessity of central data storage. The concept can be formalized as follows:

Consider \( \mathcal{D} = \{D_1, D_2, \ldots, D_n\} \) representing the distributed datasets across \( n \) clients, with \( D_i \) as the local dataset of the \( i^{th} \) client. The objective in FL is to train a global model \( \mathcal{M} \) by aggregating locally computed updates, achieved by minimizing a loss function \( \mathcal{L} \) over the model parameters \( \theta \):

\begin{equation}
\min_{\theta} \mathcal{L}(\theta) = \sum_{i=1}^{n} \frac{|D_i|}{|D|} \mathcal{L}_i(\theta)
\end{equation}

where \( |D_i| \) denotes the number of samples in \( D_i \), \( |D| \) the total number of samples across all clients, and \( \mathcal{L}_i \) the loss function computed on the \( i^{th} \) client's dataset.

Federated Averaging (FedAvg) is a prominent algorithm in FL, where each client computes an update to the model based on its local data. These updates are averaged at a central server to update the global model. The update rule in FedAvg is given by:

\begin{equation}
\theta^{(t+1)} = \theta^{(t)} - \eta \sum_{i=1}^{n} \frac{|D_i|}{|D|} \nabla \mathcal{L}_i(\theta^{(t)})
\end{equation}

Here, \( \theta^{(t)} \) and \( \theta^{(t+1)} \) represent the global model parameters before and after the update in round \( t \), respectively, and \( \eta \) is the learning rate. This aggregation process ensures that the global model learns from the entire distributed dataset while keeping each client's data localized.

Federated Learning, especially through FedAvg, enables collaborative model training in a privacy-preserving manner, suitable for scenarios where data privacy and security are critical.

\subsection{Machine Unlearning}
Machine Unlearning is the process of removing the influence of specific data points from a trained model. In scenarios where data needs to be forgotten or corrected, machine unlearning ensures that the model no longer retains or reflects this information. A common approach to machine unlearning is retraining, which involves modifying the model to reflect the absence of the data to be unlearned.

The retraining approach typically requires identifying the data that needs to be unlearned and then retraining the model from scratch or from a certain point without including this data. Formally, if a model \( \mathcal{M} \) is trained on a dataset \( \mathcal{D} \), and a subset \( \mathcal{D}_{\text{unlearn}} \subset \mathcal{D} \) needs to be unlearned, the model is retrained on the modified dataset \( \mathcal{D} \setminus \mathcal{D}_{\text{unlearn}} \). This process can be represented as:

\begin{equation}
\mathcal{M}_{\text{new}} = \text{Train}(\mathcal{M}, \mathcal{D} \setminus \mathcal{D}_{\text{unlearn}})
\end{equation}

Here, \( \mathcal{M}_{\text{new}} \) represents the updated model after retraining. The retraining process ensures that the influence of \( \mathcal{D}_{\text{unlearn}} \) is effectively removed from \( \mathcal{M}_{\text{new}} \).

While retraining is conceptually straightforward, it can be computationally intensive, especially for large models and datasets. In the context of federated learning, retraining for unlearning purposes must be coordinated across all participating clients, which adds to the complexity but is essential for maintaining the integrity and privacy compliance of the learning process.

\subsection{Blockchain}

Blockchain technology is a decentralized ledger system that provides a secure and transparent way to record transactions across multiple nodes in a network. Fundamentally, a blockchain is a chain of blocks, each containing a list of transactions. These blocks are linked using cryptographic principles, ensuring the integrity and immutability of the recorded data.

A blockchain network typically operates on the principles of consensus algorithms, which are mechanisms used to achieve agreement on a single data value among distributed processes or systems. Popular consensus algorithms in blockchain include Proof of Work (PoW) and Proof of Stake (PoS). These algorithms play a crucial role in validating transactions and maintaining the security of the blockchain.

Transactions on a blockchain are grouped into blocks, and each block is identified by a cryptographic hash of its contents, along with the hash of the previous block, thus forming a chain. This structure ensures that once data is recorded in a block, it cannot be altered without changing all subsequent blocks, which requires network consensus. This characteristic makes blockchain an ideal technology for applications that require a high degree of trust and transparency, such as smart contracts and decentralized applications (dApps).

In the context of federated learning and machine unlearning, the blockchain can provide a transparent and immutable record of machine unlearning requests and actions. This ensures that the unlearning process is not only effective, but also verifiable and auditable, addressing key challenges in ensuring data privacy and compliance in distributed learning environments.

\section{Problem Definition and System Model}

\subsection{Problem Definition}

This research addresses the challenge of integrating machine unlearning into Federated Learning (FL) systems, enhanced by Blockchain technology. The problem encompasses the following key components:

\subsubsection{Federated Learning Environment}

In a typical Federated Learning setup, a set of clients \( C = \{C_1, C_2, \ldots, C_n\} \) collaboratively train a shared model \( M \) using their local datasets \( D = \{D_1, D_2, \ldots, D_n\} \). This training occurs without the actual exchange of data, reflecting the privacy-preserving nature of Federated Learning. Each client \( C_i \) contributes to the global model by training a local model \( M_i \) on its dataset \( D_i \), and these local models are aggregated to form the global model \( M \). The challenge lies in coordinating this collaborative training process while ensuring data privacy and model efficacy.

\subsubsection{Machine Unlearning Requirement}

Machine unlearning, in this context, refers to the process of removing the influence of specific datasets \( D_{ui} \), where \( D_{ui} \subset D_i \) for a client \( C_i \), from the trained global model \( M \). This requirement arises for reasons such as data correction, compliance with legal requests for data removal, or ethical considerations. The key challenge is ensuring that the unlearning process is both effective and verifiable. It is crucial that once a dataset \( D_{ui} \) is unlearned, its influence is completely and demonstrably eliminated from the global model, a task complicated by the distributed nature of Federated Learning.

\subsubsection{Blockchain Integration}

The introduction of a decentralized ledger \( L \) into the federated learning system presents a novel solution to the challenges of machine unlearning. This ledger records and verifies all unlearning requests and actions, denoted as \( R_{ui} \) for dataset \( D_{ui} \). By leveraging Blockchain's inherent characteristics of immutability and transparency, we can ensure that each step in the unlearning process is permanently recorded and openly verifiable. This integration aims to maintain the integrity of the unlearning process, making it tamper-proof and transparent to all participants in the Federated Learning system.

\subsubsection{Trust and Verification}

Ensuring trust and verifiability in the machine unlearning process is imperative. The model must be designed to allow transparent verification that, once a data point \( D_{ui} \) is unlearned, its influence is completely absent from the model \( M \). This requirement raises significant challenges in creating mechanisms for such verification that do not compromise data privacy or the integrity of the federated learning process. The solution must balance the need for transparency with the fundamental requirements of data security and privacy intrinsic to federated learning.

In the following section, we will elaborate on the system model that addresses these challenges, describing the integrated operation of federated learning, machine unlearning, and blockchain technologies within our proposed framework.

\subsection{System Model}

Our system model innovatively integrates machine unlearning with blockchain technology, employing specialized clients and smart contracts to ensure an efficient and transparent management of the unlearning process within a secure blockchain network. This model includes a machine unlearning mechanism, two types of clients for training and unlearning, smart contracts for process automation, and a blockchain network for secure, immutable record-keeping.

\subsubsection{Clients}

The system employs two types of clients:

\begin{itemize}
    \item \textbf{Training Clients:} These clients manage routine training processes, including local model training on datasets and contributing to the overall learning and updating of the global model.
    
    \item \textbf{Unlearning Clients:} Dedicated to handling unlearning requests, these clients identify the data points \(D_{ui}\) to be unlearned and submit these requests. They are pivotal in ensuring regulatory compliance and maintaining privacy standards.

\end{itemize}

\subsubsection{Smart Contracts}

Smart Contracts play a crucial role in automating the Machine Unlearning process. Triggered by unlearning requests from unlearning clients, these self-executing contracts on the Blockchain network activate the Machine Unlearning mechanism. They meticulously log each unlearning request and action, adhering to predefined rules and conditions, thus introducing a level of trust and automation to the system.

\subsubsection{Blockchain Network}

The Blockchain network forms the foundation of our system's secure and transparent record-keeping. It chronicles all machine unlearning activities, encompassing requests from unlearning clients and the corresponding actions executed by Smart Contracts. The network's immutable and transparent nature guarantees a tamper-proof and verifiable record, augmenting the security and integrity of the machine unlearning process.

In summary, our system model is a comprehensive amalgamation of machine unlearning, specialized clients, smart contract automation, and blockchain technology. It adeptly tackles the challenges of managing the data unlearning process in a transparent and secure manner, ensuring the integrity and compliance of learning models in diverse applications.

\subsubsection{Machine Unlearning Mechanism}

Central to our system is the machine unlearning machanism, tasked with the precise removal of designated data points \( D_{ui} \) from the trained model \( M \). Initiated by smart contracts, this mechanism ensures the thorough and efficient execution of the unlearning process, preserving the integrity and performance of the model after data removal.

\section{Proposed System}

\subsection{Overview}

Our research introduces an innovative system that integrates machine unlearning with blockchain technology, aiming to enhance the trustworthiness, transparency, and compliance of learning models in various applications. This system is centered around a machine unlearning mechanism, which is tasked with the precise and efficient removal of specific data points \( D_{ui} \) from the trained model \( M \), while maintaining the model's integrity and performance after unlearning.

A key feature of our system's architecture is the strategic integration of blockchain technology. This technology serves as a secure, transparent, and immutable platform for documenting all activities related to machine unlearning, encompassing unlearning requests, actions implemented, and their outcomes. The use of blockchain ensures that every phase of the unlearning process is permanently and verifiably recorded, thereby significantly boosting the system's accountability.

\begin{figure}
\includegraphics[width=0.5\textwidth]{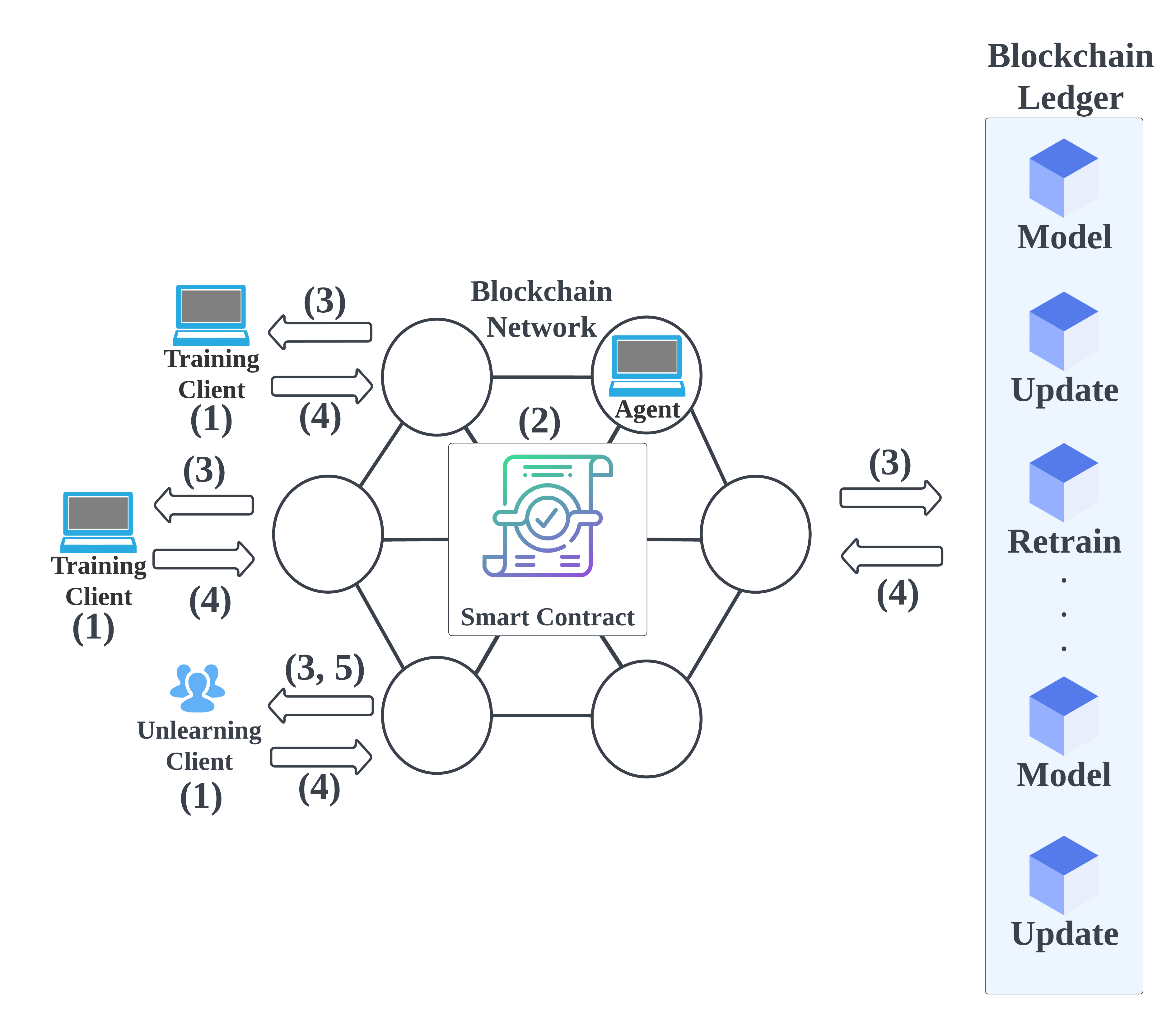}
\caption{Overview and process of our proposed system. (1) Client and agent register. (2) Global model update. (3) Model updating process. (4) Model aggregation process. (5) Machine unlearning in blockchain network}
\label{system}
\end{figure}

Moreover, the system utilizes specialized clients for distinct operational roles. Training clients are responsible for the regular training and updating of the model, whereas unlearning clients exclusively manage the initiation of unlearning requests. These clients work in coordination with the blockchain's smart contracts, which automate and regulate the machine unlearning process according to predefined protocols, thereby enhancing the process's consistency and dependability. The overview of our proposed system is shown in Figure~\ref{system}.

\subsection{Implementation of our designed system}

\subsubsection{Register}
In our system, every \textit{Client} and \textit{Agent} is required to enroll in the blockchain network for participation. The registration process, outlined in Algorithm \ref{alg:1}, is essential for integrating both \textit{Client} and \textit{Agent} into the network securely.

\begin{algorithm}
\caption{Client and Agent Register}
\label{alg:1}
\begin{algorithmic}[1]
\REQUIRE \textit{C$_{id}$, \textit{Agent}}
\ENSURE \textit{RegisterSucess, jwt}
\STATE \textit{RegisterSucess = False};
\IF{\textit{C$_{id}$ $\in$ U$_{pool}$}}
    \RETURN \textit{C$_{id}$} already existed.
\ENDIF
\IF{\textit{A$_{id}$ $\in$ U$_{pool}$}}
    \RETURN \textit{A$_{id}$} already existed.
\ENDIF
\STATE ($P_k$, $S_k$) $\xleftarrow{}$\textit{keyGenerator()};
\STATE \textit{jwt} = generateJWT($P_k$, $S_k$);
\STATE \textit{C$_{id}$, \textit{Agent}} $\xleftarrow{}$ $jwt$ $\xleftarrow{}$ SC;
\STATE \( U_{\text{pool}} = U_{\text{pool}} \cup \{C_{\text{id}}, A_{\text{id}}\} \);
\STATE \textit{RegisterSuccess = True};
\RETURN \textit{RegisterSucess, jwt}
\end{algorithmic}
\end{algorithm}

In the proposed framework, the Algorithm~\ref{alg:1} plays a pivotal role in the secure registration of clients and agents. Initially, the algorithm sets the registration success to false and checks whether the client (\(C_{id}\)) and agent (\(A_{id}\)) identifiers already exist in the user pool (\(U_{pool}\)). If an identifier is already present, the algorithm terminates, indicating duplication. For new identifiers, it employs a \textit{keyGenerator} function to create a unique pair of keys (\(P_k\) and \(S_k\)), which are then used to generate a JSON Web Token (jwt). This jwt is crucial for secure information transmission and identity verification. Subsequently, the algorithm updates the user pool with the new identifiers, marking the successful registration of the entities. It concludes by returning the registration status and the generated jwt, ensuring a robust foundation for the secure operation of the system. This algorithmic approach effectively mitigates duplicate entries and maintains the integrity of the registration process, which is essential for the system's overall functionality.


\subsubsection{Global Model Upload}
Upon successful registration within the blockchain network, the \textit{Agent} is enabled to upload the global model. Detailed in Algorithm \ref{alg:2-modified}, the Global Model Upload procedure facilitates the Agent in transferring a global model to the network. This algorithm requires inputs such as the Agent's \textit{jwt} token and the global model (\textit{$Model_g$}) intended for upload. Its outputs include an \textit{UploadModel} success indicator and the global model (\textit{Model}) that has been uploaded.

\begin{algorithm}
\caption{Global Model Upload}
\label{alg:2-modified}
\begin{algorithmic}[1]
\REQUIRE \( A_{\text{id}}, \text{jwt}, Model \)
\ENSURE \textit{UploadSuccess}, \( Model_g \)

\STATE \( \textit{UploadSuccess} = \text{False} \)

\IF{\textit{jwt token ineligibility}}
    \RETURN \text{Agent jwt token expired}
\ENDIF

\STATE \textit{Agent} sends \( Model \) to SC

\STATE \textit{SC} verifies and uploads the global model \( Model \) to the blockchain network

\STATE \( Model_g \) = \( Model \)

\STATE \( \textit{UploadSuccess} = \text{True} \)

\RETURN \( \textit{UploadSuccess}, Model_g \)
\end{algorithmic}
\end{algorithm}

Algorithm~\ref{alg:2-modified} delineates the process for agents to upload models onto a blockchain network. This procedure commences with the algorithm requiring the agent's identifier (\(A_{\text{id}}\)), a valid JSON Web Token (jwt), and the model intended for upload. Initially, the upload success is marked as false. The algorithm then verifies the jwt token's eligibility; if the token is expired or invalid, the process is halted with a notification of the jwt token's expiration. Upon successful verification, the agent transmits the model to the Smart Contract (SC), which in turn validates and uploads the model to the blockchain network. This model is then designated as the global model (\(Model_g\)). Following this, the algorithm updates the upload success status to true, indicating the successful completion of the upload process. The algorithm concludes by returning both the upload success status and the global model, thus ensuring the secure and verified integration of models into the blockchain network.


\subsubsection{Model Updating Process}
Following the setup of the global model upload, it's crucial to configure the epoch settings for the federated learning training process. Algorithm \ref{alg:3}, named "Model Updating Process," is designed to set up these parameters, specifically focusing on the epoch and batch size required for model training. The algorithm takes the epoch and batch size as input parameters and outputs a success indicator, \textit{ModelUpdate}, along with the epoch parameter.

\begin{algorithm}
\caption{Model Updating Process}
\label{alg:3}
\begin{algorithmic}[1]
\REQUIRE \textit{$epoch$, $batch size$}

\ENSURE \textit{gradient}


\STATE \textit{ModelUpdate} = False;
\STATE Clients get the \textit{$Model_g$} and use for training process;
\STATE Set \textit{$epoch$} and \textit{$batch size$} to SC
\FOR{$epoch = 1$ to $n$}

\STATE Client upload the \textit{gradient} to SC;
\IF{\textit{Client jwt token ineligibility}}
    \RETURN \textit{Client jwt token} expired
    \ENDIF
\STATE SC add DP to \textit{gradient} to achieve \textit{DP(gradient)};
\STATE SC publish the \textit{DP(gradient)} in blockchain network;
\ENDFOR
\STATE \textit{ModelUpdate = True};
\RETURN \textit{ModelUpdate}
\end{algorithmic}
\end{algorithm}

The process begins with setting the \textit{ModelUpdate} flag to false. Clients then retrieve the global model (\textit{$Model_g$}) for their training processes. The algorithm proceeds to establish the \textit{$epoch$} and \textit{$batch size$} parameters in the Smart Contract (SC). The \textit{epoch} setting will be adjusted to enhance system efficiency. During the training, for each epoch from 1 to n, clients upload their computed gradients to the SC. The algorithm also incorporates a check for the validity of the client's jwt token; if the token is found to be invalid, it returns a notification of the token's expiration and terminates. For each valid epoch, the SC adds \textit{DP()} to received gradients and then publishes the DP(gradients) on the blockchain network. Once all epochs are processed, the \textit{ModelUpdate} flag is set to true, indicating successful completion of the model updating process. This algorithm plays a vital role in ensuring the synchronization and consistency of training across clients, contributing to the overall accuracy and effectiveness of the federated learning system.

\subsubsection{Model Aggregation Process}
The model aggregation process detailed in Algorithm \ref{alg:4}, is a crucial component of the proposed system, focusing on the aggregation of model updates. It requires a JWT token and gradient inputs to perform model aggregation, indicated by \textit{ModelAggregation}, and to update the global model (\(Model_g\)).

\begin{algorithm}
\caption{Model Aggregation Process}
\label{alg:4}
\begin{algorithmic}[1]
\REQUIRE \textit{jwt token}, \textit{$gradient$}

\ENSURE \textit{ModelAggregation, $Model_g$}


\STATE \textit{ModelAggregation} = False;
\STATE \textit{Agent} get \textit{gradient} for blockchain network;
\IF{\textit{jwt token ineligibility}}
    \RETURN \textit{Agent jwt token} expired
    \ENDIF
\STATE \textit{Model} $\xleftarrow{}$ \textit{$Model_g$};
\STATE Agent update the \textit{Model} and send the new \textit{Model} to SC;
\STATE SC upload the \textit{Model} to blockchain network;
\STATE \textit{$Model_g$} $\xleftarrow{}$ \textit{Model} ;

\STATE \textit{ModelAggregation = True};
\RETURN \textit{ModelAggregation, $Model_g$}
\end{algorithmic}
\end{algorithm}

The procedure commences with setting the \textit{ModelAggregation} flag to false. The agent, after retrieving gradients from the blockchain network, undergoes a check for the JWT token's validity. An expired token leads to the termination of the process with an indication of the agent's JWT token expiration.

Once the token is verified as valid, the agent updates the model using the obtained gradients. This updated model is then sent to the Smart Contract (SC), which is responsible for uploading the model to the blockchain network. The algorithm then assigns this updated version to \(Model_g\), signifying the new global model.

The process concludes with the algorithm setting the \textit{ModelAggregation} flag to true, indicating the successful aggregation of the model. The output of the algorithm is the status of \textit{ModelAggregation} and the updated global model (\(Model_g\)), playing a vital role in integrating individual model updates into a cohesive global model and enhancing the system's overall learning efficacy.

\subsubsection{Machine Unlearning in Blockchain Network}
Algorithm \ref{alg:5} is designed to facilitate the removal of personal data from a trained model in a blockchain network. This algorithm requires the client's identifier (\(C_{id}\)) and ensures the successful execution of the unlearning process, denoted as \textit{Unlearning}.

\begin{algorithm}

\caption{Machine Unlearning in Blockchain Network}
\label{alg:5}
\begin{algorithmic}[1]
\REQUIRE \textit{$C_{id}$}
\ENSURE \textit{Unlearning}
\STATE \textit{Unlearning} = False;
\STATE Client wants to do machine unlearning to unlearn personal data;
\STATE A machine unlearning request sent to SC;
\IF{\textit{jwt token ineligibility}}
    \RETURN \textit{$C_{id}$ jwt token} expired
    \ENDIF
\STATE SC find the first \textit{epoch} before this client upload the \textit{gradient};
\STATE Agent update the \textit{Model} according to SC's requirement and upload the new model \textit{Model} to SC;
\STATE \textit{Model} $\xleftarrow{}$ \textit{$Model_g$};
\STATE SC send the new model \textit{$Model_g$} to other clients;
\STATE Other clients continue the training process according to the new model \textit{$Model_g$};
\RETURN \textit{Unlearning} = True;

\end{algorithmic}
\end{algorithm}

The process begins by setting the \textit{Unlearning} flag to false. The client, intending to execute machine unlearning to remove personal data, sends a request to the Smart Contract (SC). A key step in this process is the verification of the client's JWT token; if the token is expired or invalid, the algorithm returns a message indicating the expiration of the \(C_{id}\)'s JWT token and halts the process.

Once the token's validity is confirmed, the SC identifies the first epoch before the client uploaded their gradient. Following this, the agent is tasked with updating the model based on SC's requirements. The updated model is then uploaded back to the SC. This updated model replaces the global model (\(Model_g\)), and the SC subsequently disseminates the new global model to other clients.

Other clients in the network continue their training process using this updated model, ensuring that the unlearned data is no longer part of the training process. The algorithm concludes by setting the \textit{Unlearning} flag to true, indicating the successful completion of the machine unlearning process.

This algorithm is pivotal in maintaining the privacy and security of personal data within a blockchain-based federated learning environment, allowing for the dynamic removal of data from the trained model as required.

\subsection{Case Study: Smart Healthcare Monitoring System Using Blockchain-Enhanced Federated Learning}
\subsubsection{Background}
In the context of a smart healthcare monitoring system, an application of IoT, federated learning is employed to develop collaborative models across various devices. This system faces a critical need for machine unlearning, particularly when patients revoke consent or request data corrections.

\subsubsection{Implementation of the System}
The implementation of the system, in conjunction with our proposed method, is delineated as follows:
\begin{itemize}
\item \textbf{Registration:} Patients and healthcare providers, acting as clients and agents in the system, enroll through the blockchain network. This secure registration process, detailed in Algorithm 1, is vital for their participation and guarantees data integrity.
\item \textbf{Global Model Upload:} Following registration, healthcare providers (agents) securely upload the global model to the blockchain network, as described in Algorithm 2. This step ensures that all participating devices are synchronized with the most current and accurate model.

\item \textbf{Model Updating Process:} Continuously collecting patients' health data necessitates regular model updates. Algorithm 3 defines the parameters like epoch and batch size for the training process, maintaining the model's relevance with the latest data. Additionally, Secure Computing (SC) incorporates Differential Privacy (DP) automatically, reinforcing the system's security and privacy.

\item \textbf{Model Aggregation Process:} Individual model updates from diverse IoT devices are consolidated into an updated global model, as outlined in Algorithm 4. This aggregation is crucial for a unified and comprehensive understanding of patient health trends across the network.

\item \textbf{Machine Unlearning in Blockchain Network:} Algorithm 5 enables the removal of a patient's personal data from the trained model when they opt to withdraw their data. This step is imperative for adhering to privacy requests and maintaining the system's trustworthiness.

\end{itemize}
\subsubsection{Analysis}
The integration of blockchain not only facilitated secure and verifiable data unlearning but also preserved the overall system's efficiency. The healthcare monitoring system adeptly adjusted to changes in patient data while safeguarding data privacy and maintaining system integrity.

\subsubsection{Challenges and Solutions}
The primary challenges involved optimizing the efficiency of the blockchain network and balancing security with privacy. These challenges were tackled by optimizing the federated learning training process's epoch aggregation settings and implementing Differential Privacy (DP) methods to secure the updating data.

\subsubsection{Conclusion}
This case study exemplifies the efficacy of integrating blockchain with federated learning in a practical IoT application. Our proposed framework significantly enhances data privacy, upholds the integrity of the learning model, and provides a verifiable, efficient mechanism for machine unlearning, which is paramount in sensitive domains like healthcare monitoring.

\section{Privacy and security analysis}

\subsection{Privacy analysis}
Our innovative framework integrating machine unlearning with blockchain in a federated learning context significantly bolsters data privacy. This section delves into the privacy benefits and mechanisms of our system.

\textbf{Data Localization in Federated Learning:} A cornerstone of our framework is the localization of data on client devices. This approach inherently minimizes privacy risks commonly associated with centralized data storage. By keeping sensitive data on the client side and only sharing model updates, the framework upholds privacy by design.

\textbf{Machine Unlearning for Privacy Enhancement:} Machine unlearning is pivotal for privacy protection in our system. It allows for the precise and complete removal of data from the model upon user request or policy changes. This ensures that any data, once deemed unnecessary or sensitive, is effectively erased from the learning model, safeguarding user privacy.

\textbf{Blockchain as a Catalyst for Privacy:} Blockchain technology plays a crucial role in enhancing privacy in our framework. Its immutable ledger records all data transactions (including unlearning requests and actions) permanently. This transparency offers a robust audit trail, which is essential for accountability, yet it does not compromise data privacy, as the ledger only records transactional metadata, not the data itself.

\textbf{Smart Contracts for Privacy Assurance:} Our system employs smart contracts to automate and enforce privacy policies. These contracts guarantee that unlearning requests are executed efficiently and accurately, thereby maintaining the system's privacy integrity. Automation reduces the risk of human errors and biases, further strengthening privacy safeguards.

\textbf{Addressing Privacy in IoT Contexts:} The IoT environment, characterized by the vast and varied data from numerous devices, presents unique privacy challenges. Our framework is designed to manage these challenges by processing IoT-generated data in a manner that prioritizes user privacy. It ensures that any data removal or unlearning request is thoroughly and promptly implemented.

In summary, our proposed system markedly advances privacy protection in federated learning settings. By synergizing federated learning, machine unlearning, and blockchain, the framework adeptly navigates the complex landscape of data privacy, particularly in IoT scenarios. It sets a new standard for future developments in balancing sophisticated data processing with the paramount need to protect individual privacy.

\subsection{Security analysis}
In addition to enhancing privacy, our integrated system of machine unlearning with blockchain in a federated learning environment also significantly strengthens security. This section examines the key security aspects of our proposed framework.

\textbf{Robust Data Security in Federated Learning:} The federated learning aspect of our system ensures that data remains decentralized, greatly reducing the risk of large-scale data breaches common in centralized systems. By distributing the data across multiple nodes, the system inherently disperses security risks, making it more resilient to targeted attacks.

\textbf{Immutable Record-Keeping with Blockchain:} The implementation of blockchain provides an additional layer of security. Its immutable ledger ensures that every transaction, including the addition and removal of data, is permanently recorded. This immutability makes it virtually impossible to tamper with the data history, ensuring the integrity of the entire learning model.

\textbf{Smart Contract-Driven Security Protocols:} Our system leverages smart contracts to enforce security protocols automatically. These contracts are programmed to execute unlearning requests securely and ensure compliance with predefined security standards. This automation minimizes human intervention, thereby reducing the potential for errors and security lapses.

\textbf{Enhanced Security through Machine Unlearning:} Machine unlearning contributes to the system's security by ensuring that any data no longer needed or deemed sensitive can be removed promptly and completely. This not only protects against unauthorized access to outdated or irrelevant data but also reduces the 'attack surface' that hackers could exploit.

\textbf{Security Challenges in IoT Environments:} The diversity and scale of IoT networks present unique security challenges. Our framework addresses these by ensuring secure data handling and model training across various devices. The blockchain component adds a trust layer, verifying and recording each action taken within the network, thus safeguarding against malicious activities.

\textbf{Resilience Against Data Tampering and Attacks:} The combination of blockchain technology and federated learning creates a formidable barrier against data tampering and cyber-attacks. Blockchain’s distributed nature makes it difficult for attackers to compromise the system, while federated learning limits the exposure of sensitive data.

In conclusion, our proposed framework presents a comprehensive approach to enhancing security in federated learning environments. By integrating machine unlearning and blockchain technology, it addresses critical security concerns, particularly relevant in the context of IoT. This framework not only ensures the secure handling of data but also builds trust among participants, paving the way for more secure and reliable data-driven applications.

\section{Performance evaluation}

In this section, we provide a thorough evaluation of our system's performance, focusing on two key aspects: the effectiveness of the machine unlearning process within federated learning environments, and the efficiency and dependability of blockchain integration. To assess our system, we employed two renowned datasets, MNIST and CIFAR-10, to test it across various learning scenarios. Additionally, we analyzed the complexity and performance effects of incorporating blockchain technology into the federated learning framework. The detailed results and analyses of our experiments are presented in Sections VII-A and VII-B.

\subsection{Experimental Configuration}

Our research experiments were executed on a high-performance hardware setup to guarantee both accuracy and reliability of the results. The cornerstone of our hardware infrastructure was an Intel(R) Xeon(R) Gold 6238R CPU, running at 2.20GHz. This processor was chosen for its powerful computational capabilities, essential for handling the intensive data processing demands of our experiments. Alongside the CPU, we utilized 86 GB of LPDDR4 memory, which provided substantial capacity and speed to manage large datasets and intricate algorithms efficiently. Furthermore, a Quadro RTX 5000 GPU was integrated into our system, enhancing our machine learning tasks with its advanced capabilities. This entire hardware setup was operated on the RED HAT 7.9 operating system, chosen for its stability and high performance in demanding computing environments.

On the software front, our experiments leveraged Hyperledger Fabric Version 2.X, a robust platform ideal for implementing and managing our system’s blockchain component. Hyperledger Fabric was selected due to its scalability, security features, and support for complex blockchain applications. For programming the blockchain network, we used the Go programming language, recognized for its efficiency and suitability in building scalable network applications. The development and execution of the code were performed in Visual Studio Code, an integrated development environment (IDE) offering a flexible and user-friendly platform for programming. To thoroughly test the effectiveness of our proposed system, we employed two widely recognized datasets: MNIST and CIFAR-10. These datasets are standard benchmarks within the machine learning community, allowing us to assess our system’s performance under realistic and challenging conditions.

\subsection{Results and Analysis}

\subsubsection{MNIST Dataset Results }
The MNIST dataset, comprising handwritten digits, served as our primary benchmark for evaluation. This dataset, known for its simplicity and clear learning tasks, was ideal for assessing the accuracy and efficiency of our machine unlearning process. We specifically focused on how unlearning requests affected the model's accuracy and training efficiency, observing the system's behavior under various volumes and frequencies of these requests.

\begin{figure*}[h]
  \centering
  \begin{minipage}[b]{0.19\textwidth}
    \centering
    \includegraphics[width=\textwidth]{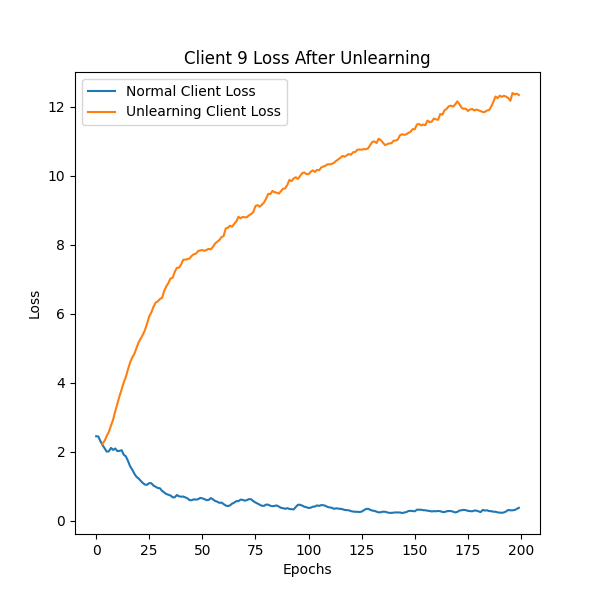}
    \caption{Specific Client loss comparison in 200 epochs}
    \label{m1}
  \end{minipage}
  \begin{minipage}[b]{0.19\textwidth}
    \centering
    \includegraphics[width=\textwidth]{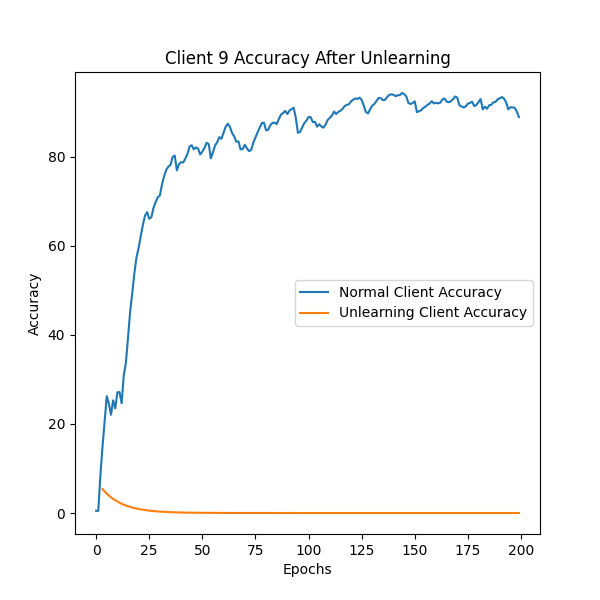}
    \caption{Specific Client accuracy comparison in 200 epochs}
    \label{m2}
  \end{minipage}
  \begin{minipage}[b]{0.19\textwidth}
    \centering
    \includegraphics[width=\textwidth]{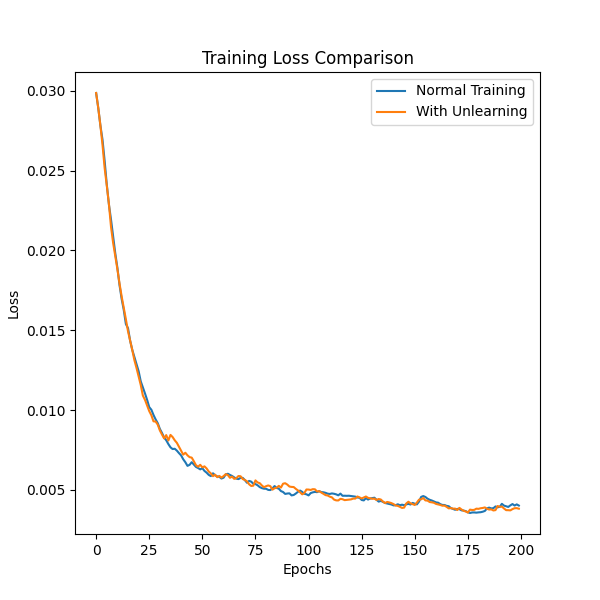}
    \caption{Loss comparison of normal learning and with unlearning in 200 epochs}
    \label{m3}
  \end{minipage}
  \begin{minipage}[b]{0.19\textwidth}
    \centering
    \includegraphics[width=\textwidth]{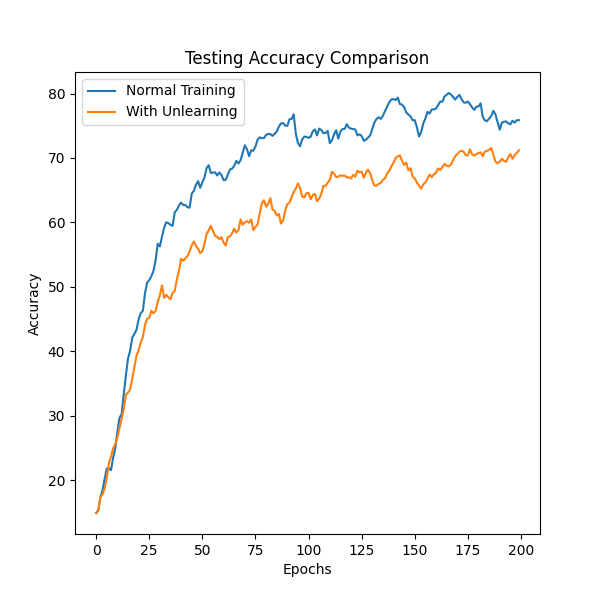}
    \caption{Testing accuracy comparison of normal learning and with unlearning in 200 epochs}
    \label{m4}
  \end{minipage}
  \begin{minipage}[b]{0.19\textwidth}
    \centering
    \includegraphics[width=\textwidth]{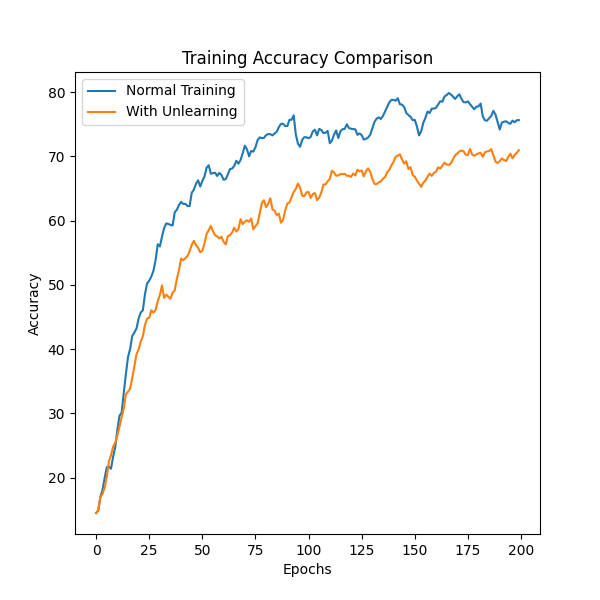}
    \caption{Training accuracy comparison of normal learning and with unlearning in 200 epochs}
    \label{m5}
  \end{minipage}

  \centering
  \begin{minipage}[b]{0.19\textwidth}
    \centering
    \includegraphics[width=\textwidth]{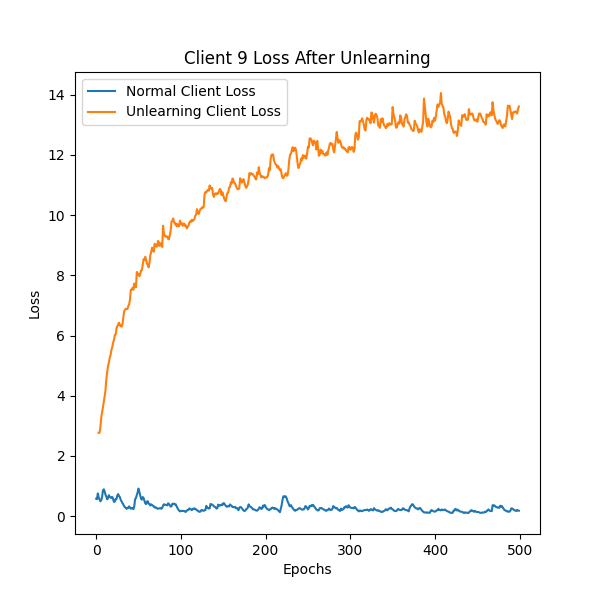}
    \caption{Specific Client loss comparison in 500 epochs}
    \label{m7}
  \end{minipage}
  \begin{minipage}[b]{0.19\textwidth}
    \centering
    \includegraphics[width=\textwidth]{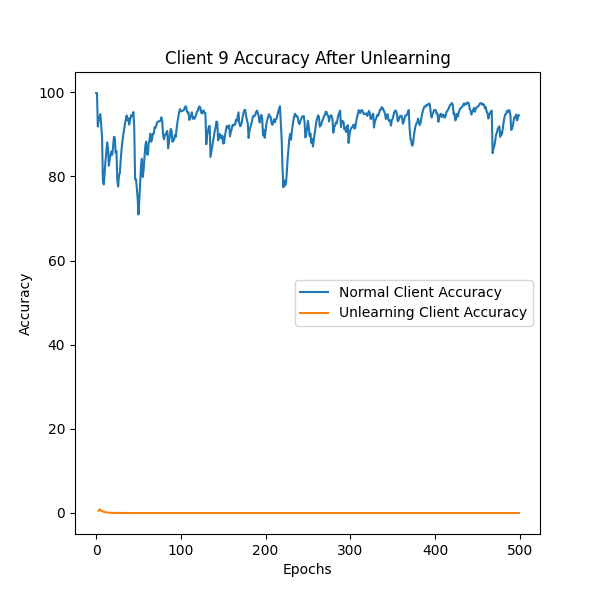}
    \caption{Specific Client accuracy comparison in 500 epochs}
    \label{m8}
  \end{minipage}
  \begin{minipage}[b]{0.19\textwidth}
    \centering
    \includegraphics[width=\textwidth]{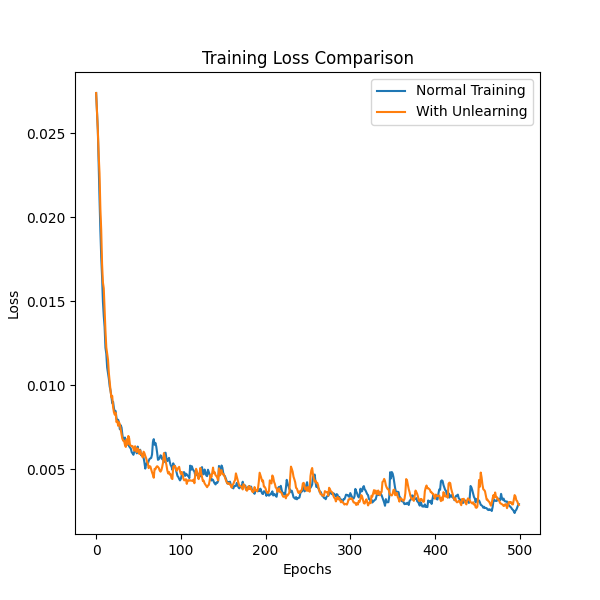}
    \caption{Loss comparison of normal learning and with unlearning in 500 epochs}
    \label{m9}
  \end{minipage}
  \begin{minipage}[b]{0.19\textwidth}
    \centering
    \includegraphics[width=\textwidth]{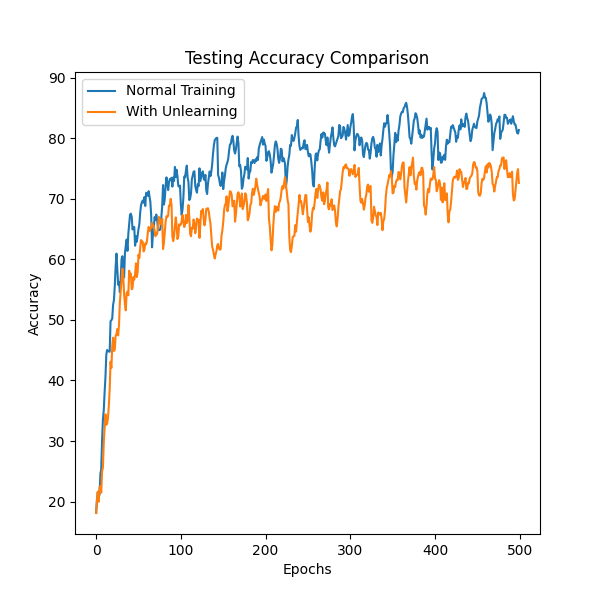}
    \caption{Testing accuracy comparison of normal learning and with unlearning in 500 epochs}
    \label{m10}
  \end{minipage}
  \begin{minipage}[b]{0.19\textwidth}
    \centering
    \includegraphics[width=\textwidth]{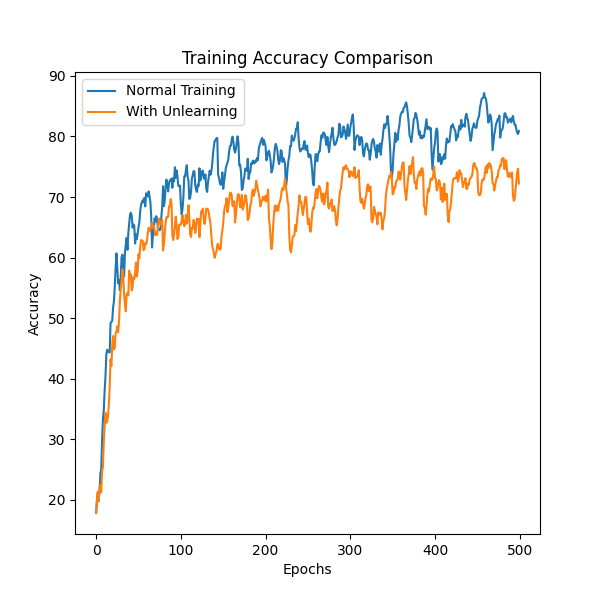}
    \caption{Training accuracy comparison of normal learning and with unlearning in 500 epochs}
    \label{m11}
  \end{minipage}
\end{figure*}

In evaluating the MNIST dataset's performance, we closely analyzed the model's response during and after the unlearning process. As illustrated in Figure~\ref{m1}, we compared loss trends between normal training and post-unlearning phases. During the training epochs, the loss trajectory for normal training consistently remained lower than that of the unlearning phase. This pattern began with a sharp decline in the early epochs, followed by stabilization over time. Notably, while there was a temporary increase in loss for the unlearning client, this eventually leveled off, approximating the normal client loss, albeit at a slightly higher value.

Figure~\ref{m2} compares the accuracy during normal and unlearning phases for the same client. The accuracy in normal training quickly reached a high plateau, with minor fluctuations, indicative of a stable and effective learning process. In contrast, the accuracy for the client post-unlearning started at zero and remained constant, reflecting the targeted removal of class information from the model.

\begin{figure}[htbp]
\centering
    \begin{minipage}[b]{0.24\textwidth}
    \centering
    \includegraphics[width=\textwidth]{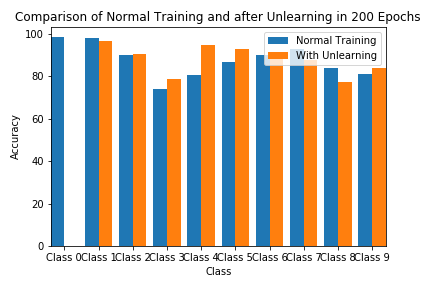}
    \caption{Class comparison in 200 epochs}
    \label{m6}
  \end{minipage}
  \begin{minipage}[b]{0.24\textwidth}
    \centering
    \includegraphics[width=\textwidth]{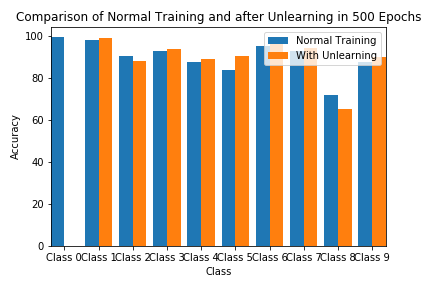}
    \caption{Class comparison in 500 epochs}
    \label{m12}
  \end{minipage}
\end{figure}

The overall training loss, depicted in Figure~\ref{m3}, highlighted the system's resilience. Both normal training and training with unlearning converged to a loss value close to zero, indicating that unlearning did not compromise the model's ability to learn from the remaining dataset. Similarly, Figures~\ref{m4} and \ref{m5} displayed minimal difference in accuracy between normal training and training with unlearning. Both scenarios showed a sharp initial increase in accuracy, followed by closely aligned progression, suggesting that the unlearning process did not significantly affect the overall model accuracy.
The most definitive proof of our system's unlearning accuracy is presented in Figure~\ref{m6}. Here, the accuracy for Class $0$ post-unlearning was precisely zero, confirming the complete and effective removal of this class from the model's knowledge. For normal training, the average accuracy across all classes was 80.23\%, while it slightly reduced to $71.35\%$ post-unlearning. This decrease was mainly due to the targeted unlearning in Class 0, as the accuracies for other classes remained largely unaffected, demonstrating the specificity of the unlearning process.

These results underscore our system's ability to perform machine unlearning with high specificity and minimal impact on overall model performance. The precision of the unlearning process, along with the retained efficacy for the remaining classes, emphasizes the system's applicability in real-world machine learning scenarios where data privacy and the right to be forgotten are critical considerations.

Extending our evaluation to $500$ epochs on the MNIST dataset, we derived insightful conclusions:

Figures~\ref{m7} and \ref{m8}, representing 'Client 9 Loss' and 'Client 9 Accuracy' post-unlearning, demonstrate the success of the unlearning process. In Figure~\ref{m7}, the normal client loss remains low and stable, contrasting with the unlearning client loss, which, after initial stability, gradually increases. This divergence distinctly illustrates the impact of unlearning over extended training.

Figure~\ref{m8} shows a significant difference in accuracy. While the normal client's accuracy consistently exceeds $80\%$, indicating stable model performance over time, the unlearning client's accuracy, initially parallel, plummets to near zero. This stark decline exemplifies the effective elimination of the learned data points.

From a broader view, Figures~\ref{m9} and ~\ref{m10}, showcasing 'Training Loss Comparison' and 'Testing Accuracy Comparison,' reveal a slight divergence between normal training and training with unlearning. Both figures indicate a marginally increased loss and reduced accuracy in training with unlearning, particularly in later stages. This suggests a cumulative effect of unlearning over time. Nonetheless, the overall loss remains low and the accuracy high, emphasizing the system's resilience.

Figure~\ref{m11} reinforces the observation of sustained high accuracy levels throughout the epochs. The unlearning curve, while slightly lower in accuracy, aligns with the testing accuracy trends.
The most evident unlearning effects are depicted in Figure~\ref{m12}. Normal training maintains a high accuracy of $87.11\%$, whereas post-unlearning accuracy drops to $77.74\%$. This decrease is most notable in Class 0 accuracy, which falls to zero, confirming the targeted and successful unlearning. The minor decrease in accuracy for other classes, likely due to the unlearning of Class 0, does not significantly impact the overall model performance.

In conclusion, these findings validate the precision of our unlearning process within a federated learning context. The system skillfully unlearns specific data points without undermining the model's overall learning capacity, essential for maintaining privacy and compliance with data regulations. The slight overall accuracy reduction post-unlearning highlights the unlearning impact while affirming the system's capability to maintain robust performance across the remaining dataset.

\subsubsection{CIFAR-10 Dataset Results}

The CIFAR-10 dataset, comprising $60,000$ color images across 10 classes, provided a more complex test environment. Utilizing this dataset, we evaluated the system's ability to handle intricate image data and assessed the scalability of our machine unlearning process. We focused on examining the model's resilience and adaptability to unlearning requests, while maintaining accuracy and learning efficiency in this intricate scenario.

Upon extending our experiments to include the CIFAR-10 dataset over $2000$ epochs, we observed several key dynamics in model performance:
\begin{figure*}[htbp]
  \centering
  \begin{minipage}[b]{0.19\textwidth}
    \centering
    \includegraphics[width=\textwidth]{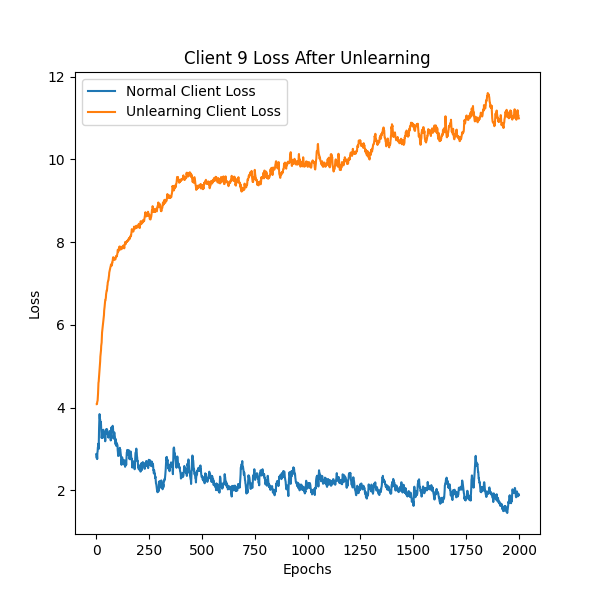}
    \caption{Specific Client loss comparison in 2000 epochs}
    \label{c1}
  \end{minipage}
  \begin{minipage}[b]{0.19\textwidth}
    \centering
    \includegraphics[width=\textwidth]{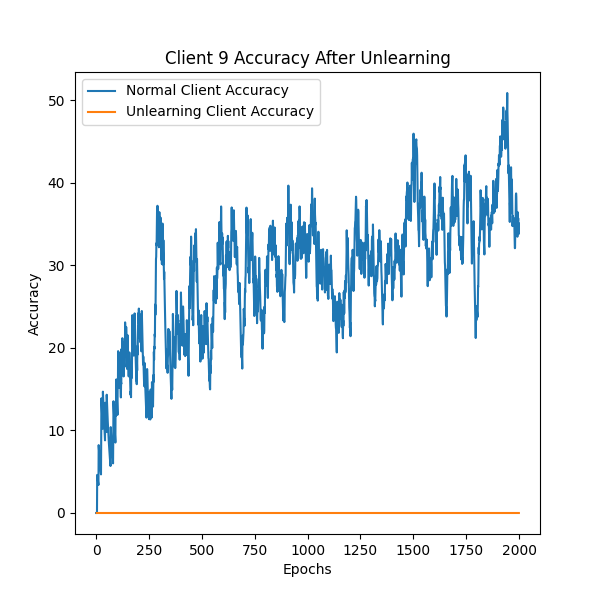}
    \caption{Specific Client accuracy comparison in 2000 epochs}
    \label{c2}
  \end{minipage}
  \begin{minipage}[b]{0.19\textwidth}
    \centering
    \includegraphics[width=\textwidth]{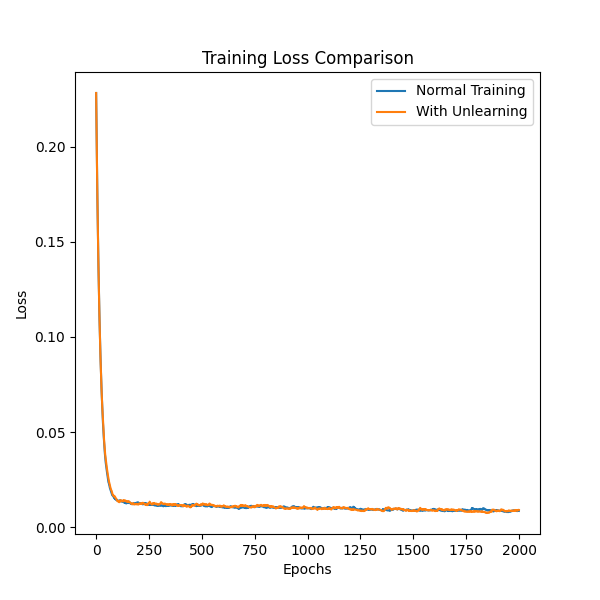}
    \caption{Loss comparison of normal learning and with unlearning in 2000 epochs}
    \label{c3}
  \end{minipage}
  \begin{minipage}[b]{0.19\textwidth}
    \centering
    \includegraphics[width=\textwidth]{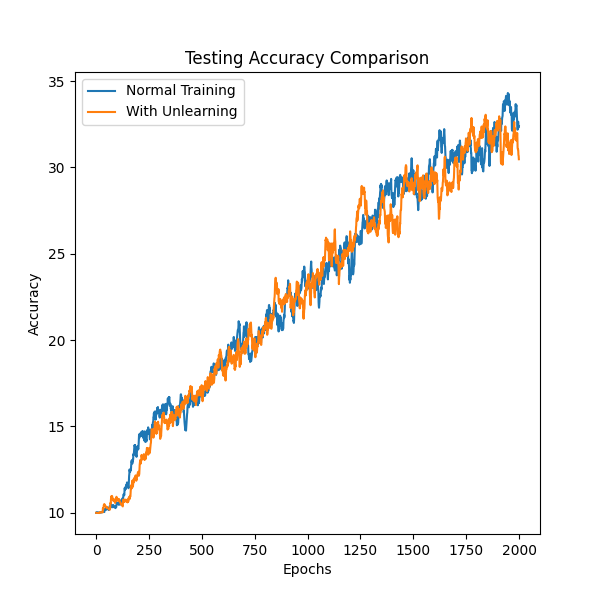}
    \caption{Testing accuracy comparison of normal learning and with unlearning in 2000 epochs}
    \label{c4}
  \end{minipage}
  \begin{minipage}[b]{0.19\textwidth}
    \centering
    \includegraphics[width=\textwidth]{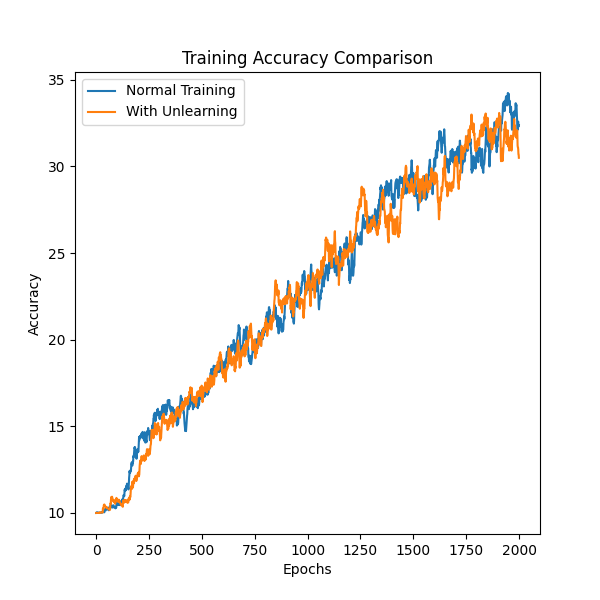}
    \caption{Training accuracy comparison of normal learning and with unlearning in 2000 epochs}
    \label{c5}
  \end{minipage}

  \centering
  \begin{minipage}[b]{0.19\textwidth}
    \centering
    \includegraphics[width=\textwidth]{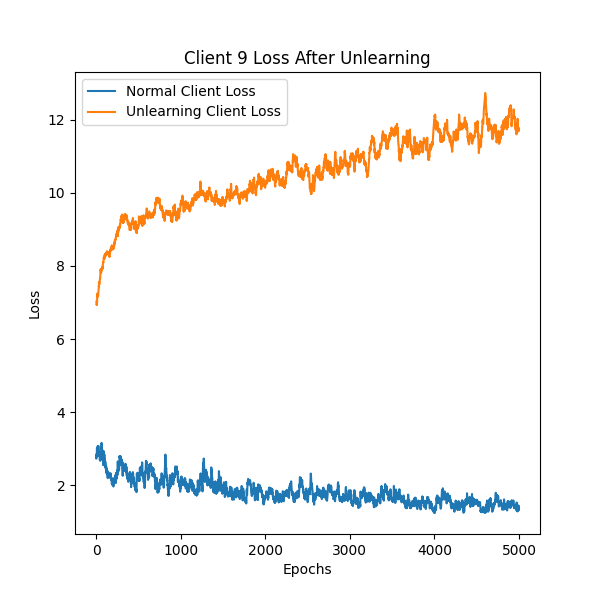}
    \caption{Specific Client loss comparison in 5000 epochs}
    \label{c7}
  \end{minipage}
  \begin{minipage}[b]{0.19\textwidth}
    \centering
    \includegraphics[width=\textwidth]{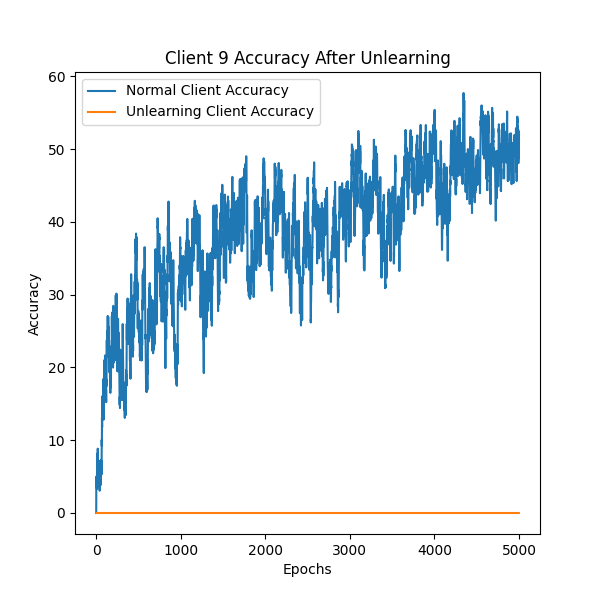}
    \caption{Specific Client accuracy comparison in 5000 epochs}
    \label{c8}
  \end{minipage}
  \begin{minipage}[b]{0.19\textwidth}
    \centering
    \includegraphics[width=\textwidth]{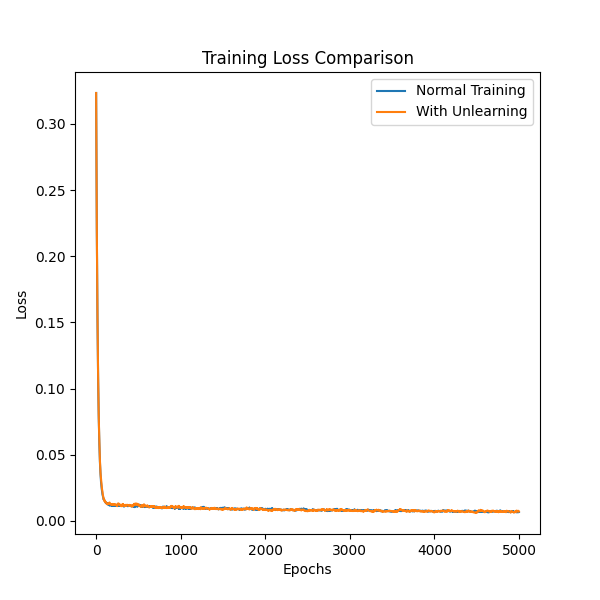}
    \caption{Loss comparison of normal learning and with unlearning in 5000 epochs}
    \label{c9}
  \end{minipage}
  \begin{minipage}[b]{0.19\textwidth}
    \centering
    \includegraphics[width=\textwidth]{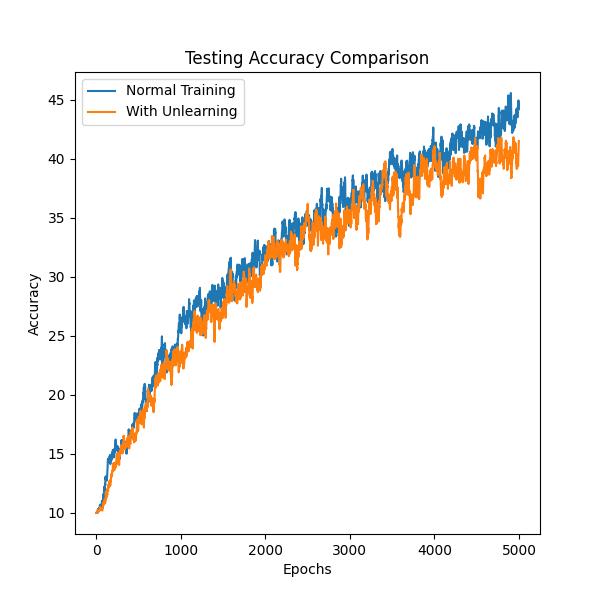}
    \caption{Testing accuracy comparison of normal learning and with unlearning in 5000 epochs}
    \label{c10}
  \end{minipage}
  \begin{minipage}[b]{0.19\textwidth}
    \centering
    \includegraphics[width=\textwidth]{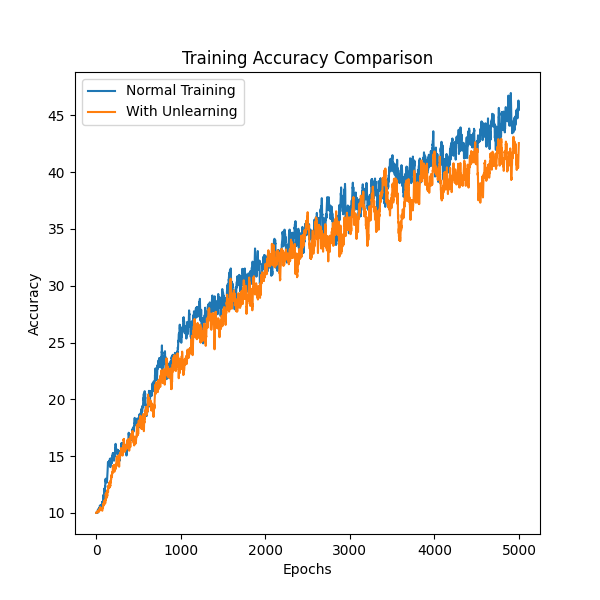}
    \caption{Training accuracy comparison of normal learning and with unlearning in 5000 epochs}
    \label{c11}
  \end{minipage}
\end{figure*}

Figure~\ref{c1} illustrates the loss trajectory for the normal client, beginning with a rapid decline indicative of effective initial learning, then stabilizing at a lower value. In contrast, the unlearning client exhibited a moderate increase in loss over time. This pattern suggests that, despite a slight performance degradation, the unlearning process does not inhibit the client's ongoing learning capacity.

In Figure~\ref{c2}, a significant disparity in accuracy is evident. The normal client's accuracy fluctuates around $40\%$, reflecting the typical behaviour of a converging learning model. Conversely, the unlearning client’s accuracy quickly falls to nearly zero and remains consistently low, highlighting the unlearning mechanism's effectiveness.

Figure~\ref{c3} shows that the system maintains its learning ability post-unlearning, with both normal training and training with unlearning converging to a minimal loss. Similarly, Figure~\ref{c4} indicates that while accuracy in the unlearning scenario is slightly reduced, it tracks closely with normal training, implying robust overall model performance.

\begin{figure}[htbp]
\centering
    \begin{minipage}[b]{0.24\textwidth}
    \centering
    \includegraphics[width=\textwidth]{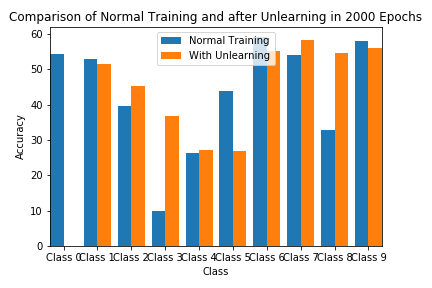}
    \caption{Class comparison in 2000 epochs}
    \label{c6}
  \end{minipage}
  \begin{minipage}[b]{0.24\textwidth}
    \centering
    \includegraphics[width=\textwidth]{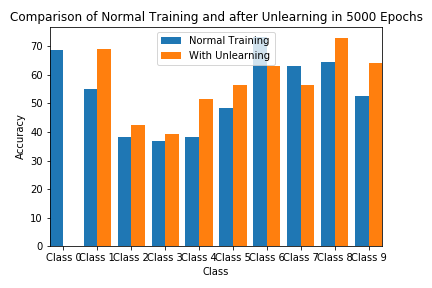}
    \caption{Class comparison in 5000 epochs}
    \label{c12}
  \end{minipage}
\end{figure}

Figure~\ref{c5} supports these observations, displaying a consistent gap between normal training and training with unlearning. Notably, the model trained with unlearning still achieves high accuracy, validating the unlearning process's precision.
The class-wise accuracy comparison in Figure~\ref{c6} offers compelling evidence of the unlearning process's impact. Post-unlearning, the accuracy for Class 0 drops to zero, showing the model's complete inability to recognize the unlearned class, clearly evidencing the success of the unlearning process. The average accuracy across all classes slightly decreases from $34.12\%$ to $33.21\%$ post-unlearning, with the most significant reduction in the targeted class. This outcome illustrates the system's ability to selectively unlearn specific information while retaining general knowledge, an essential aspect for practical applications requiring selective data removal.

In conclusion, the evaluation using the CIFAR-10 dataset over an extended timeframe highlights the system's effectiveness in executing precise machine unlearning. The marginal overall accuracy decline post-unlearning underscores the unlearning process's targeted nature, ensuring that the model retains its learning capacity and maintains the integrity of its performance across the remaining classes.

Extending our CIFAR-10 dataset analysis over $5000$ epochs, we gained valuable insights into the model's adaptability and the unlearning process's efficacy:

In Figure~\ref{c7}, we observed the loss patterns for the normal and unlearning clients. The normal client's loss showed a steep initial decline, signalling effective learning, and then stabilized at a low level, indicating a consistent learning pattern. Conversely, the unlearning client's loss followed a similar initial trend but gradually increased, eventually stabilizing at a higher level than the normal client. This pattern suggests that, although the unlearning client continues to learn, the unlearning process introduces a slight but noticeable alteration in its loss trajectory.

Figure~\ref{c8} illustrates the accuracy of the normal client, which fluctuates within an acceptable range and peaks around $50\%$. For the unlearning client, the accuracy initially matches the normal client but then rapidly declines to zero. This dramatic drop demonstrates the successful elimination of the influence of the unlearned data from the model.

In Figure~\ref{c9}, the trajectories for both normal training and training with unlearning initially exhibit a sharp decline. However, as the epochs progress, the curves converge closely, suggesting that the unlearning process does not significantly impair the model's ability to learn from the remaining data.

Figure~\ref{c10} shows the accuracy trends for normal training and training with unlearning. Both follow a parallel ascent, with the latter maintaining a slightly lower accuracy level. This indicates that the model continues to generalize effectively post-unlearning, albeit with a minor decrease in accuracy.

Figure~\ref{c11} aligns with the testing accuracy trends, where training with unlearning shows a consistent but modest divergence from normal training. The slight decrease in accuracy after unlearning, though evident, does not significantly detract from the model's performance, highlighting the unlearning process's precision.
The class-wise accuracy comparison in Figure~\ref{c12} starkly demonstrates the unlearning impact. While normal training achieves an average accuracy of 45.92\%, post-unlearning accuracy drops to $42.65\%$. Notably, the accuracy for Class 0 is zero post-unlearning, confirming the effective and specific execution of the unlearning. The other classes show only minor reductions in accuracy, consistent with the unlearning's targeted nature.

Overall, the extended analysis over $5000$ epochs confirms the system's effectiveness in executing machine unlearning. The approach significantly reduces the model's accuracy for the specified unlearned class while minimally impacting the overall performance. This capability underscores our system's utility in scenarios where data privacy and regulatory compliance are crucial, facilitating selective unlearning as needed.

\subsubsection{Blockchain Complexing Results}

In our study, we thoroughly examined the blockchain component of our system, focusing on aspects such as scalability, transaction throughput, and latency due to blockchain integration in the federated learning process. Our objective was to assess the impact of these factors on the overall system performance, ensuring that blockchain integration does not compromise the efficiency or effectiveness of the machine unlearning process.

We utilized Hyperledger Fabric 2.X to evaluate the impact of the blockchain network on our machine unlearning process's performance, particularly in IoT applications where additional computational overhead is a concern.

    
    
    
\begin{itemize}
    \item \textbf{Blockchain Network Initialization}: The initial setup time for the blockchain network was approximately 35 seconds. This one-time overhead is deemed acceptable, especially considering the long-term benefits in federated learning applications where rapid deployment is essential.

    \item \textbf{Consensus Mechanism Overhead}: The time required for the consensus process varied based on the chosen consensus algorithm. After setting up the blockchain network, our consensus algorithm, requiring approval from all nodes, added around 3 seconds. This duration is reasonable and manageable within our IoT context involving federated learning.

    \item \textbf{Transaction Processing Efficiency}: The average time for processing transactions, including model updates and gradient aggregation, was 2 seconds. This efficiency in transaction handling by Hyperledger Fabric highlights its capability to manage additional blockchain-related tasks effectively.

    \item \textbf{Per-Epoch Time Cost}: During training, the duration per epoch, both for normal training and post-unlearning operations, remained consistent at 24-26 seconds, illustrating the system's stable performance irrespective of unlearning activities.
\end{itemize}

Table~\ref{table:time} presents a comparison of time costs between a standard federated learning cycle and our proposed blockchain-enhanced method. Initially, our method incurs a higher time cost due to setup and endorsement processes. However, this cost tends to normalize with increasing iterations, indicating promising scalability.

\begin{table}[h]
\centering
\caption{Time Cost Analysis for Federated Learning with and without Blockchain Integration over 1999 Iterations}
\label{table:time}
\begin{tabular}{|p{3.5cm}|c|c|c|c|}
\hline
\textbf{Method} & \textbf{t = 0} & \textbf{t = 9} & \textbf{t = 199} & \textbf{t = 1999}\\ \hline
Normal Federated Learning & 26s & 260s & 5200s & 26000s\\ \hline
Our Proposed System & 66s & 318s & 5638s & 28038s\\ \hline
\end{tabular}
\end{table}

Overall, our results confirm that blockchain technology can be practically and scalably integrated into federated learning frameworks with unlearning capabilities. The additional time cost is balanced by the enhanced security and trustworthiness in the machine unlearning process, highlighting the suitability of our approach for IoT applications where these features are paramount.

\subsubsection{Analysis Conclusion}
Drawing on the empirical results from the MNIST and CIFAR-10 datasets, along with the blockchain time complexity analysis, we can draw a cohesive conclusion:

Our system, which integrates machine unlearning into federated learning environments and supplements it with blockchain technology, has proven to be both effective and efficient. The evaluation using the MNIST dataset demonstrates that the unlearning process is executed with precision, having minimal impact on the overall model performance while ensuring the targeted data's removal. This finding is supported by the results from the CIFAR-10 dataset, showing that even in more complex scenarios, the system successfully unlearns specific classes without significantly affecting the accuracy of other classes.

In extended trials over thousands of epochs with the CIFAR-10 dataset, the system consistently exhibited its capability to execute machine unlearning precisely while maintaining high model accuracy. The slight reduction in accuracy post-unlearning highlights the targeted impact of the unlearning process, critical in scenarios prioritizing data privacy and the right to be forgotten.

From the perspective of blockchain complexity, our study utilized the Hyperledger Fabric 2.X platform to evaluate the impact of blockchain integration on machine unlearning performance. Our analysis indicated that the initial time overhead is offset by the blockchain's contributions to security and trustworthiness. The blockchain's scalability is evident, as demonstrated by the normalization of time costs with increased iterations, reinforcing the system's suitability for long-term applications.

The blockchain time complexity results confirm that the blockchain network's overhead is manageable and does not significantly impact the federated learning process. The initial setup time for the blockchain network and the latency due to the consensus mechanism are acceptable, particularly given the enhanced security and immutability benefits that blockchain brings.

In summary, our proposed system emerges as a scalable, secure, and efficient solution for federated learning, particularly in IoT scenarios. It effectively addresses the critical challenge of integrating machine unlearning without sacrificing learning performance or scalability. The inclusion of a blockchain layer augments trust and security, making our approach highly relevant for applications where these aspects are crucial. The comprehensive analyses across different datasets, coupled with an in-depth study of blockchain time complexity, collectively affirm the practicality and viability of our blockchain-enhanced federated learning system.

\section{Conclusion} 

In conclusion, our research presents a pioneering approach that synergizes machine unlearning with blockchain technology within the realm of federated learning, particularly tailored for Internet of Things (IoT) applications. The extensive experimental analysis conducted on the MNIST and CIFAR-10 datasets has successfully demonstrated the efficiency of our system in performing machine unlearning with minimal impact on the overall model performance. This underscores the effectiveness of our methodology in balancing precision in unlearning with maintaining overall model accuracy.

Additionally, the time complexity analysis of our blockchain component, implemented using Hyperledger Fabric 2.X, has highlighted the practicality of integrating blockchain into federated learning systems. Despite introducing a reasonable overhead, the incorporation of blockchain enhances the security and trustworthiness of the system, crucial for IoT applications.

Looking to the future, our research trajectory includes several ambitious goals: enhancing the efficiency of the blockchain component, extending our approach to more complex datasets, exploring advanced consensus mechanisms for blockchain, integrating differential privacy techniques for enhanced data security, probing into cross-chain interoperability to broaden the scope of our system, and evaluating the system's adaptability to ever-evolving data regulations. These endeavours aim to continuously refine and adapt our system, ensuring its applicability and relevance in the dynamic domain of secure, distributed machine learning in IoT and federated learning applications. Through sustained research and development efforts, we aspire to keep pace with the rapid advancements in this field and contribute to its growth and innovation.

\end{document}